\documentclass[
    ,final            % use final for the camera ready runs
  ]
  {aipproc}

\usepackage{amsmath}
\layoutstyle{8x11double}

\begin{document}

\title{Recent HERA Results on Leptoquarks and other SUSY-related Signatures}

\classification{
%<Replace this text with PACS numbers; choose from this list:
%                \texttt{http://www.aip..org/pacs/index.html}>
12.15.-y,
12.60.-i,
12.60.Jv,
13.38.Be,
13.60.Hb,
13.85.Fb,
13.85.Hd,
13.85.Qk,
13.85.Rm,
13.90.+i,
14.80.-j,
14.80.Ly
}

\author{Stefan Schmitt}{
  address={DESY, Notkestr.~85, 22607 Hamburg, GERMANY}
}

\begin{abstract}
  The HERA $ep$ collider and the experiments H1 and
  ZEUS operated from 1994-2007. A total integrated
  luminosity of almost $1\,\text{fb}^{-1}$ was collected at
  centre-of-mass energies up to 320~GeV. Results from searches for
  leptoquarks and squarks, final states with an isolated lepton and
  missing transverse momentum and final states with multi-leptons are
  presented. The leptoquark limits are interpreted in terms of limits
  on squark production in SUSY models with $R$-parity violating couplings.
\end{abstract}

\maketitle

%%%%%%%%%%%%%%%%%%%%%%%%%%%%%%%%%%%%%%%%%%%%
%% MAINMATTER
%%%%%%%%%%%%%%%%%%%%%%%%%%%%%%%%%%%%%%%%%%%%

\section{Searches for Leptoquarks at HERA}

At HERA, leptoquarks (LQs) may be resonantly produced by the fusion of the
incoming lepton and a quark originating from the proton. Depending on
their quantum numbers and couplings, these objects may decay into a
charged or neutral lepton and a quark.
However, for both reactions $ep\to eX$ and $ep\to \nu X$ there is
irreducible background from deep-inelastic scattering. 
For LQ masses below the kinematic limit of HERA ($320\,\text{GeV}$) one may be able to see LQs as a resonant
structure in the reconstructed mass $M_{LQ}$. If the LQ mass is higher,
LQs may show up as contact interactions.
\begin{figure}
  \includegraphics[width=0.35\textwidth]{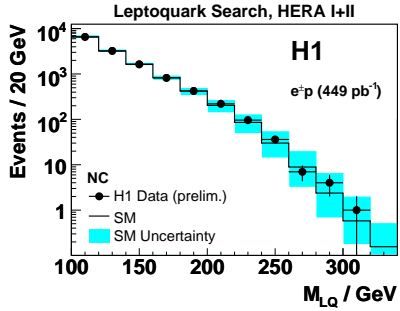}
  \caption{\label{figure:h1lqmassnc}H1 reconstructed mass spectra for $LQ\to eq$.}
\end{figure}
\begin{figure}
  \includegraphics[width=0.35\textwidth]{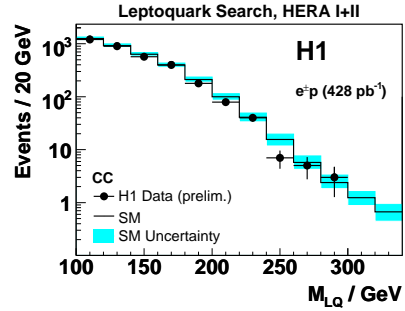}
  \caption{\label{figure:h1lqmasscc}H1 reconstructed mass spectra for $LQ\to \nu q$.}
\end{figure}

\begin{figure}
  \includegraphics[width=0.35\textwidth]{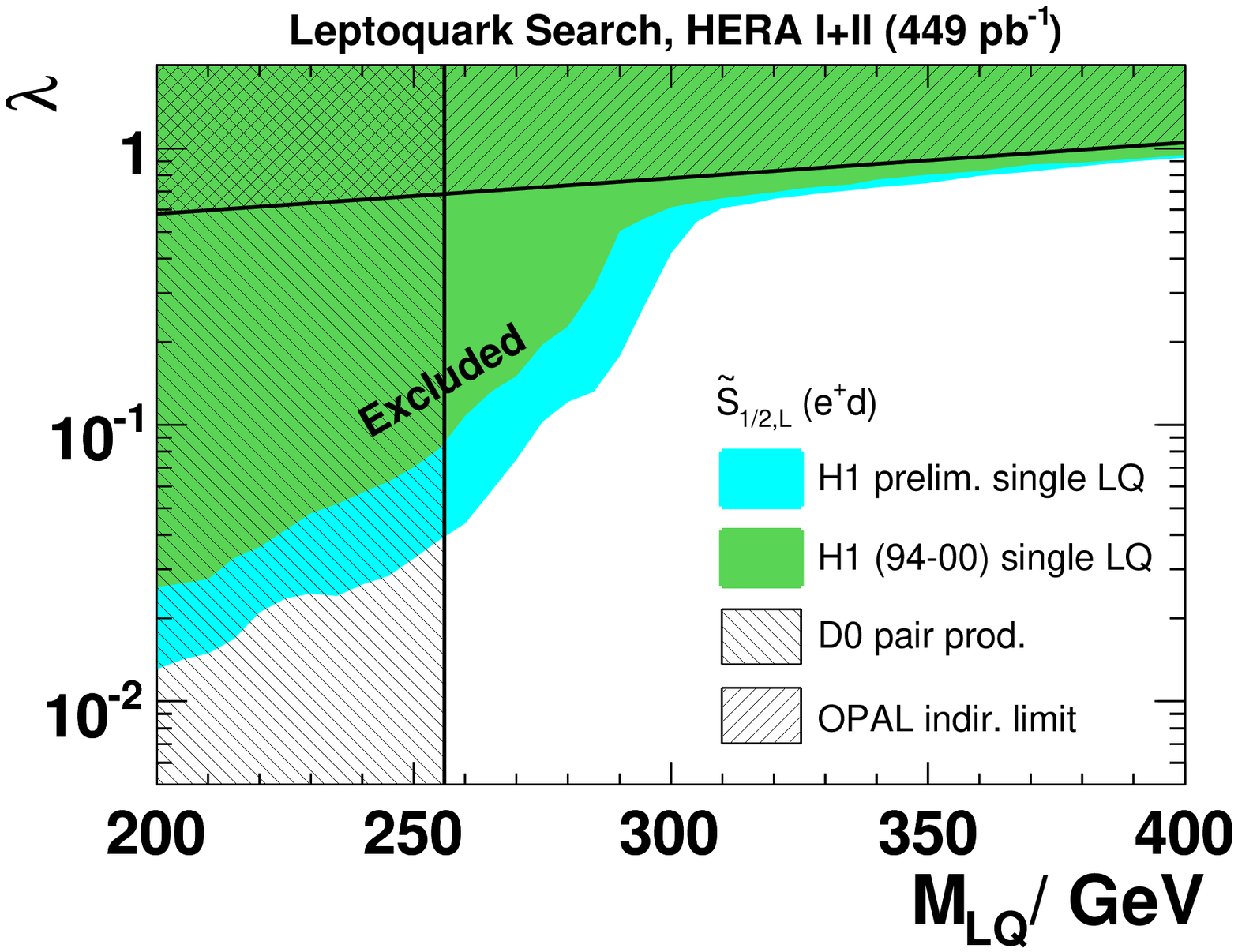}
  \includegraphics[width=0.35\textwidth]{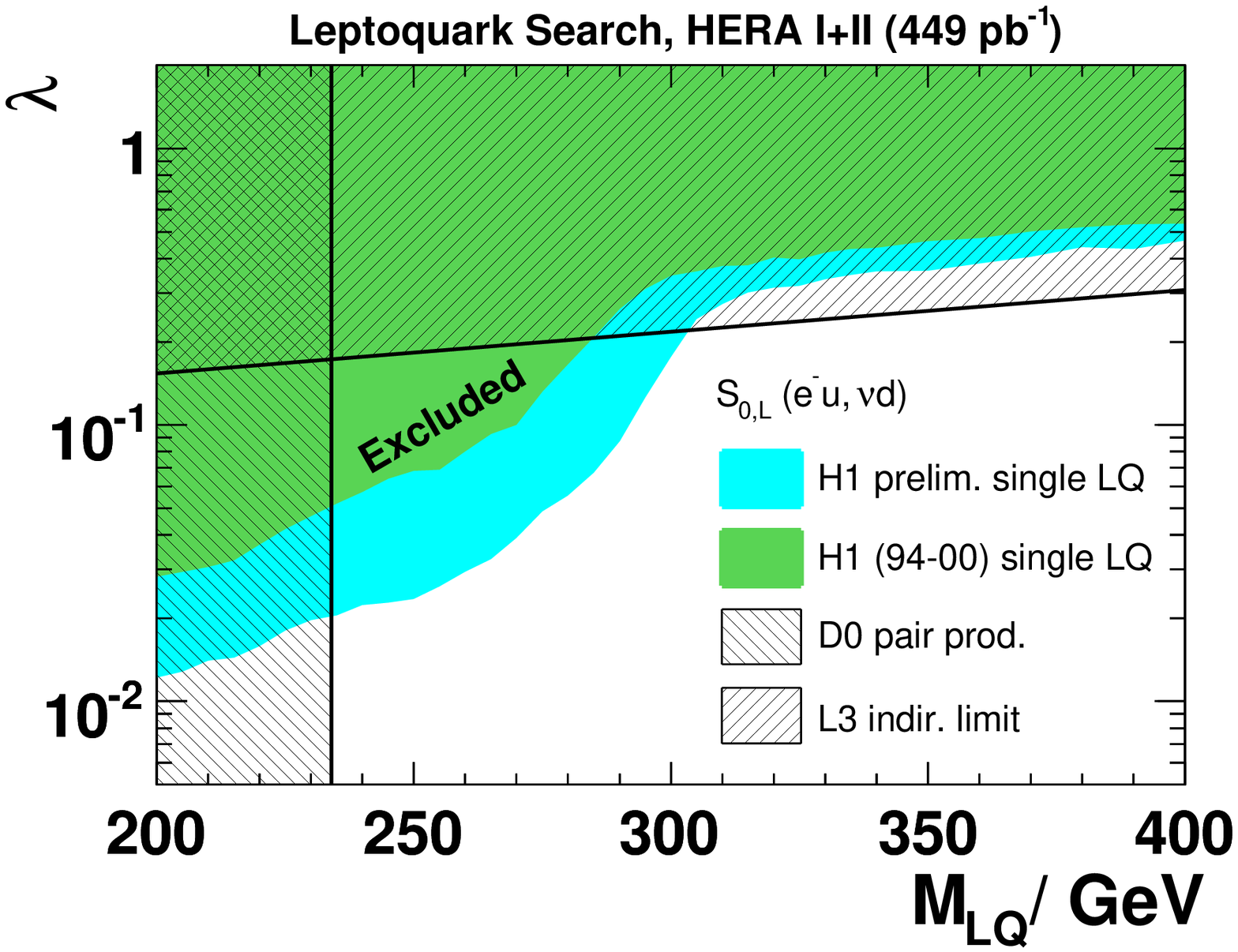}
  \caption{\label{figure:h1lqlimits}H1 preliminary $95\%$ confidence limits
    on the production of $\tilde{S}_{1/2L}$ and $S_{0L}$
    leptoquarks, corresponding to $\tilde{u}_i$ and $\tilde{d}_j$
    squarks in $R_P$ violating models, respectively. Limits on the coupling
    $\lambda$ are shown as a function of the leptoquark mass.}
\end{figure}

Dedicated searches for LQs and contact interactions mediated
by LQs have been reported earlier
\cite{Adloff:2003jm,Chekanov:2003pw,Chekanov:2003af,Aktas:2005pr} for 
a subset of the HERA data, collected from 1994-2000. 
Here updated results, comprising the full HERA datasets are presented \cite{h1prelimlq}.
Figure~\ref{figure:h1lqmassnc} and \ref{figure:h1lqmasscc} show the
reconstructed LQ mass spectra seen by the H1 collaboration.
No evidence for LQ production is seen, and limits are set on
the LQ coupling $\lambda$ as a function of the LQ mass. A total of 14
LQ models are investigated \cite{Buchmuller:1986zs}. Two LQs, namely the
$\tilde{S}_{1/2,L}$ and the $S_{0,L}$ are of special interest, because
these can be interpreted as a squark produced via $R$-parity ($R_P$) violating
coupling $\lambda^{\prime}_{1ij}$ \cite{Butterworth:1992tc}. The
correspondence of LQ type and SUSY particle is indicated in
Table~\ref{table:zeuslqci}. Limits on LQ production at $95\%$
confidence 
level, determined from the complete H1 data are shown in Figure
\ref{figure:h1lqlimits}. They may be interpreted as squark limits for 
SUSY models where the direct $R_P$ violating decay dominates or
for models with squark masses larger than the HERA centre-of-mass energy. 
For couplings of electromagnetic strength, $\lambda=0.3$, the
production of $\tilde{S}_{1/2L}$ and $S_{0L}$ LQs are excluded for
masses up to $295$ GeV and $310$ GeV, respectively.

The single-differential cross-sections $d\sigma/dQ^2$ in
neutral-current deep-inelastic scattering are sensitive to LQ 
production, mediated by contact interactions. Again, 
$\tilde{S}_{1/2L}$ and $S_{0L}$ LQs may be interpreted as squarks in
$R_P$ violating SUSY models. The ZEUS collaboration
reports preliminary limits on the ratio of LQ mass to coupling
$M/\lambda$, using $274\,\text{pb}^{-1}$ of data \cite{zeusprelimci}. 
\begin{table}
\renewcommand{\arraystretch}{1.3}
\begin{tabular}{ccccc}
\tablehead{5}{c}{b}{ZEUS 1994-2005 (prel.) $e^{\pm}p$} \\
\tablehead{3}{c}{b}{Scalar leptoquarks} &\tablehead{2}{c}{b}{Vector leptoquarks} \\
\tablehead{1}{c}{b}{LQ} & \tablehead{1}{c}{b}{$\tilde{q}$} & \tablehead{1}{r}{b}{$95\%$ C.L. [TeV]} &
\tablehead{1}{c}{b}{LQ} & \tablehead{1}{c}{b}{$95\%$ C.L. [TeV]} \\
\hline
$S_0^L$ & $\tilde{d}_j$& $0.96$ & $V_0^L$ & 0.80 \\
$S_0^R$ & & $0.82$ & $V_0^R$ & 0.62 \\
$\tilde{S}_0^R$ & & $0.32$ & $\tilde{V}_0^R$ & $1.33$ \\
$S_{1/2}^L$ & $\tilde{u}_i$ & 0.88 & $V_{1/2}^L$ & 0.46 \\
$S_{1/2}^R$ & & 0.46 & $V_{1/2}^R$ & 1.00 \\
$\tilde{S}_{1/2}^L$ & & 0.44 & $\tilde{V}_{1/2}^L$ & 1.10 \\
$S_1^L$ & & $0.74$ & $V_1^L$ & 1.91 \\
\hline
\end{tabular}
\caption{ZEUS exclusion limits on $M/\lambda$ in
  contact interactions mediated by leptoquarks.
\label{table:zeuslqci}}
\end{table}
The $95\%$ confidence limits on are summarised in Table
\ref{table:zeuslqci}.

\begin{figure}
  \includegraphics[width=0.3\textwidth]{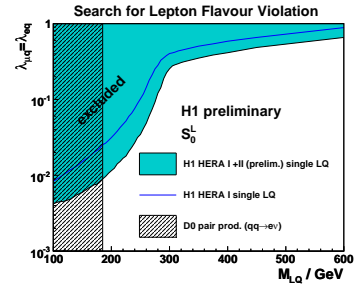}
  \caption{\label{figure:h1lfvlimit}H1 preliminary $95\%$ confidence limits
    on the production of a $S_{0L}$ leptoquark with flavour-violating
    couplings $\lambda_{\mu q'}=\lambda_{eq}$ ($\lambda_{\tau q'}=0$).}
\end{figure}

Another model for LQ production at HERA includes flavour-violating
decays. The LQ may have a couplings $\lambda_{\mu q'}$ or
$\lambda_{\tau q'}$ in addition to the coupling $\lambda_{eq}$, which
has been discussed above. At HERA such models are 
probed in the reactions $ep \to \mu X$ or $ep \to \tau X$ 
\cite{Chekanov:2005au,Aktas:2007ji}. It is worth noting that the limits on
the search for $\tilde{S}_{1/2,L}$ and $S_{0,L}$ LQs may be
interpreted as a search for squarks with off-diagonal $R_P$ violating
couplings $\lambda'_{ijk}$. Results from a search for LQs decaying to
$\mu^{-}+\text{jet}$ using the full H1 $e^{-}p$ data
\cite{h1prelimlfv} are shown in
Figure~\ref{figure:h1lfvlimit}.
For couplings of electromagnetic strength $\lambda_{\mu
  q'}=\lambda_{eq}=0.3$, assuming $\lambda_{\tau q'}=0$, the production of a
$S_{0L}$ is excluded for masses up to $305$ GeV.

\section{Signatures with isolated Lepton and missing Transverse
  Momentum}

The HERA data are searched for events with an isolated lepton ($\ell=e,\mu$) with high transverse momentum
$P_T^{\ell}>10\,\text{GeV}$ and high missing transverse momentum
$P_T^{\text{miss}}>12\,\text{GeV}$. The $P_T^{\text{miss}}$
is attributed to a neutrino which escaped detection. The main signal
process is SM real $W$ production. Heavy resonances 
would be expected to produce an excess over the SM predictions at
large values of total hadronic transverse momentum, $P_T^X$.
\begin{figure}
  \includegraphics[width=0.275\textwidth]{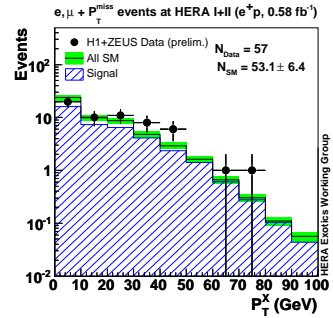}
  \caption{\label{figure:h1isolepptx}H1 and ZEUS combined data on 
    isolated leptons ($\ell=e,\mu$) with missing transverse momentum.
    The $e^{+}p$ data are shown as a function of the transverse
    momentum of the hadronic system $P_T^X$.}
\end{figure}
The data from both experiments, ZEUS and H1, are combined in a common
phase-space \cite{h1zeusprelimisolep}. Figure~\ref{figure:h1isolepptx} shows the distribution of
$P_T^X$ for the $e^{+}p$ data. At high $P_T^X$, where the SM
prediction is small, there is an excess of
events, with $23$ observed over $14.6\pm 1.9$ expected from Standard
Model (SM) processes. No such excess is present in the $e^{-}p$ data.
\begin{figure}
  \includegraphics[width=0.275\textwidth]{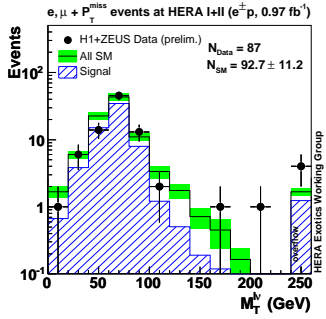}
  \caption{\label{figure:h1isolepmt}H1 and ZEUS combined data on 
    isolated leptons ($\ell=e,\mu$) with missing transverse momentum.
    The data are shown as a function of the transverse
    mass.}
\end{figure}
The transverse mass of the lepton and the neutrino is investigated in
Figure~\ref{figure:h1isolepmt}. There is clear evidence that the events are
dominated by SM single $W$ production. An excess of such events in
$e^{+}p$ data may originate in SUSY models from bosonic stop
decays, as studied in detail with HERA I data \cite{Aktas:2004tm}.
Alternatively, an enhanced production of real $W$ bosons at HERA may
originate from the decay of anomalously produced top quarks, as
previously investigated with HERA I data
\cite{Chekanov:2003yt,Aktas:2003yd}. The H1 collaboration reports a
preliminary $95\%$ confidence limit on the anomalous coupling
$\vert\kappa_{tu\gamma}\vert<0.14$, using their complete dataset \cite{h1prelimtop}.

\section{Multi-Lepton Signatures}

\begin{figure}
  \includegraphics[width=0.32\textwidth]{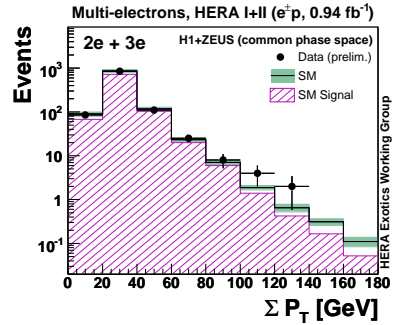}
  \caption{\label{figure:threee}H1 and ZEUS combined data on 
    multi-electrons. The data are shown as a function of the scalar
    sum of the transverse momenta.}
\end{figure}

H1 and ZEUS have searched their complete data for
signatures with two or three leptons ($\ell=e,\mu$) 
\cite{Aaron:2008jh,zeusprelimeee}. 
Such events are predicted in several exotic models, for example from
the decay of doubly charged Higgs bosons \cite{Aktas:2006nu}. The
combined H1 and ZEUS data from a search for three electron final states
\cite{h1zeusprelimeee} are shown in
Figure~\ref{figure:threee}.
At high transverse momenta $\sum P_T>100\,\text{GeV}$ there is an
upward fluctuation of 6 data events compared to $3\pm0.34$
expected. Because the HERA collider has ceased operation, it is not
possible to investigate further the nature of this fluctuation.

\section{Summary}

New results from searches for Leptoquarks, isolated leptons with
missing transverse momentum and multi-lepton signatures at HERA have been
reported. No clear evidence of new physics has been found. Compared to older
results, improved limits on Leptoquark production and anomalous top
couplings are derived.

%%%%%%%%%%%%%%%%%%%%%%%%%%%%%%%%%%%%%%%%%%%%%%%%
%% BACKMATTER
%%%%%%%%%%%%%%%%%%%%%%%%%%%%%%%%%%%%%%%%%%%%%%%%

\end{document}